\documentclass[prd, preprint, nofootinbib]{revtex4}

\usepackage{graphicx}
\usepackage{amsfonts}
\usepackage{latexsym}
\usepackage{amsmath}
\usepackage{amssymb}
\usepackage{epsfig}

\begin{document}

\vspace*{-2cm}

\title{Inelastic $B-L$ scalar dark matter and the LUX-ZEPLIN event}

\author{Nobuchika Okada}
 \email{okadan@ua.edu}
 \affiliation{
Department of Physics and Astronomy, 
University of Alabama, Tuscaloosa, Alabama 35487, USA
}

\author{Osamu Seto}
 \email{seto@particle.sci.hokudai.ac.jp}
 \affiliation{Department of Physics, Hokkaido University, Sapporo 060-0810, Japan}

%

\begin{abstract}
We show that our previously proposed inelastic scalar dark matter (DM) model in gauged $U(1)_{B-L}$ symmetry naturally accounts for the $2.6\sigma$ high-energy recoil event recently reported by the LUX-ZEPLIN (LZ) collaboration. 
To satisfy the thermal relic abundance, TeV-scale scalar DM $S$ with $m_S \simeq m_{Z'}/2$ for resonant annihilation and the gauge coupling constant $g_{B-L} \sim 0.5$ are required,
 where $m_{Z^\prime}$ is the mass of $U(1)_{B-L}$ gauge boson $Z^{\prime}$.
The resulting predicted DM-nucleon inelastic cross section of $\sigma_\mathrm{SI} \sim 10^{-45}$ cm${}^2$
 is in excellent agreement with the LZ event. 
This scenario can be decisively tested for future collider searches for $Z^\prime$ boson resonance.
\end{abstract}

\preprint{EPHOU-26-011} 

\vspace*{3cm}

\maketitle

\section{Introduction}

Weakly interacting massive particle (WIMP) is a primary candidate for the dark matter (DM)
 in the Universe. 
Among various WIMP scenarios, DM candidates in well-motivated extensions of the Standard Model (SM) are particularly appealing.
Supersymmetry (SUSY) is a prime example of a model beyond the Standard Model (BSM) and, with $R$-parity, provides a leading candidate for SUSY WIMP dark matter~\cite{Jungman:1995df}. 
A gauged $U(1)_{B-L}$ symmetry is another well-motivated extension, serving as the natural origin of neutrino masses and being embedded in Grand Unified Theories~\cite{Pati:1974yy,Davidson:1978pm,Mohapatra:1980qe,Marshak:1979fm}. 
Several years ago, we have proposed a model of inelastic scalar WIMP in gauged $U(1)_{B-L}$ model
 where light neutrino masses are induced through the seesaw mechanism with
 the Majorana right-handed neutrino masses~\cite{SeesawM,SeesawY,SeesawG,SeesawMS}
 generated by the $B-L$ symmetry breaking~\cite{Okada:2019sbb}. 
The remarkable aspect of this model is that the $Z_2$ parity to stabilize DM candidate
 is not introduced by hand but is automatically realized by the charge assignment in the model.

Attention to inelastic dark matter~\cite{Hall:1997ah,TuckerSmith:2001hy} is once again rising rapidly,
 since the release of news from the LUX-ZEPLIN (LZ) collaboration~\cite{LZ:2026axp}.
The LZ collaboration reported an anomalous single particle interaction event in their recent analysis
 of $220$ live days of data, exhibiting a significantly higher energy deposition
 than expected from standard elastic WIMP-nucleus scattering. 
This event is difficult to explain with known background signals and shows a $2.6\sigma$ deviation. 
In inelastic dark matter scenarios,
 DM particles undergo transition to a slightly heavier state by scattering, which
 naturally leads to higher energy recoil signals while it kinematically suppresses standard
 elastic scattering cross sections.
Following the announcement by the LZ collaboration, strongly degenerate Higgsino-like neutralino DM has been extensively discussed as a heavy inelastic WIMP candidate to explain the LZ event~\cite{Su:2026rwz,Fan:2026kxx,Freese:2026sga,Wu:2026nhi,Du:2026guj}
 though such scenarios may face severe astrophysical constraints and others~\cite{Pospelov:2026ewn,Rodd:2026tyn}.
For other interpretations, see for instance Refs.~\cite{Lou:2026idn,Yin:2026jnn,Nomura:2026qyq,DiMauro:2026ldr,Visinelli:2026kgt,Yamashita:2026ump,Smirnov:2026aqk,McCabe:2026crm,Jeesun:2026vzo,Unwin:2026rdp}.

In this article, we show that our previously proposed inelastic scalar DM model
 based on gauged $U(1)_{B-L}$ symmetry~\cite{Okada:2019sbb} can naturally accommodate
 the characteristics of the LZ event while satisfying the cosmological abundance as thermal relic
 and current experimental constraints.

\section{Inelastic DM model in $U(1)_{B-L}$ extended SM}
\label{Sec:Model}

\begin{table}[htbp]
\begin{center}
\begin{tabular}{|c|ccc|c|}
\hline
      &  SU(3)$_c$  & SU(2)$_L$ & U(1)$_Y$ & U(1)$_{B-L}$  \\ 
\hline
$Q^{i}$ & {\bf 3 }        &  {\bf 2}         & $ 1/6$     & $1/3 $   \\
$u^{i}_{R}$ & {\bf 3 }    &  {\bf 1}         & $ 2/3$     & $1/3 $   \\
$d^{i}_{R}$ & {\bf 3 }    &  {\bf 1}         & $-1/3$     & $1/3 $   \\
\hline
$L^{i}$        & {\bf 1 }    &  {\bf 2}       & $-1/2$     & $-1 $    \\
$e^{i}_{R}$    & {\bf 1 }    &  {\bf 1}       & $-1$       & $-1 $    \\
\hline
$\Phi$            & {\bf 1 }    &  {\bf 2}       & $ 1/2$    & $0 $   \\  
\hline
$N^{i}_{R}$    & {\bf 1 }    &  {\bf 1}       &$0$         & $-1 $    \\
$\phi_1$       & {\bf 1 }    &  {\bf 1}       &$ 0$        & $ + 1 $  \\
$\phi_2$       & {\bf 1 }    &  {\bf 1}       &$ 0$        & $ + 2 $  \\ 
\hline
\end{tabular}
\end{center}
\caption{
The particle content of our $U(1)_{B-L}$ model. 
In addition to the SM particle content ($i=1,2,3$), three RH neutrinos  
  ($N_R^i$ ($i=1, 2, 3$)) and two $U(1)_{B-L}$ Higgs fields ($\phi_1$ and $\phi_2$) are introduced.   
}
\label{tableBL}
\end{table}

We consider a model based on the gauge group
 $SU(3)_C \times SU(2)_L \times U(1)_Y \times U(1)_{B-L}$.
In addition to the SM particle content with three generations ($i=1,2,3$), three RH neutrinos  
  ($N_R^i$ ($i=1, 2, 3$)) and two $U(1)_{B-L}$ Higgs fields ($\phi_1$ and $\phi_2$) are introduced. 
With the three RH neutrinos, the model is fully free from all gauge and mixed gauge-gravitational anomalies.
The particles and those charges are listed on Table~\ref{tableBL}.
The scalar potential is expressed as~\cite{Okada:2018xdh,Chao:2017ilw}
\begin{align}
V(\Phi, \phi_1, \phi_2 )
 =& - M^2_{\Phi} |\Phi|^2 + \frac{\lambda}{2} |\Phi|^4 + M^2_{\phi_1} \phi_1\phi_1^{\dagger} - M^2_{\phi_2} \phi_2\phi_2^{\dagger}   \nonumber \\
 & + \frac{1}{2} \lambda_1 (\phi_1 \phi_1^{\dagger})^2+\frac{1}{2}\lambda_2 (\phi_2\phi_2^{\dagger} )^2
 +\lambda_3 \phi_1\phi_1^{\dagger} (\phi_2 \phi_2^{\dagger}) \nonumber \\
 & +  (\lambda_4 \phi_1\phi_1^{\dagger} + \lambda_5 \phi_2\phi_2^{\dagger})|\Phi|^2
 - A (\phi_1 \phi_1 \phi_2^{\dagger} + \phi_1^{\dagger} \phi_1^{\dagger} \phi_2 ) ,
\label{eq:totalpotential}
\end{align}
with $\Phi$ being the SM Higgs field.
All parameters in the potential~(\ref{eq:totalpotential}) are taken to be real and positive.

\subsection{Dark matter mass and interactions}

At the $U(1)_{B-L}$ and the electroweak (EW) symmetry breaking vacuum, the SM Higgs field and
 the $U(1)_{B-L}$ Higgs field are expanded around those VEVs, $v$ and $v_2$, as
\begin{align}
\Phi =& \left( \begin{array}{c}
          0 \\
          \frac{v + \varphi}{\sqrt{2}} \\
         \end{array}
  \right), \\
\phi_1 =& \frac{ S + i P }{\sqrt{2}} ,\\
\phi_2 =& \frac{ v_2 + \varphi_2 }{\sqrt{2}} .
\end{align}
With the $U(1)_{B-L}$ and the EW symmetry breaking, the $Z^\prime$ boson,
 $S$ and $P$ acquire their masses, respectively, as 
\begin{align}
  m_{Z^\prime}^2 =& 4 g_{B-L}^2 v_2^2, \\
  m_S^2 =& M_{\phi_1}^2+\frac{1}{2}\lambda_3v_2^2+\frac{1}{2}\lambda_4 v^2-\sqrt{2}A v_2, \\
  m_P^2 =& M_{\phi_1}^2+\frac{1}{2}\lambda_3v_2^2+\frac{1}{2}\lambda_4 v^2+\sqrt{2}A v_2 ,
\end{align}
 where $g_{B-L}$ is the $U(1)_{B-L}$ gauge coupling.
The parameter $A$ controls the mass splitting between $S$ and $P$. 
Since we take $A$ positive, $S$ is lighter than $P$ and becomes the DM candidate.
The parameter $A$ will be estimated, once the mass spliting is measured,
from the formula of the mass spliting
\begin{align}
  m_P^2-m_S^2 = 2\sqrt{2}A v_2 .
\end{align}
The degenerate mass between $S$ and $P$ is natural in our framework, as both states originate from the same complex scalar multiplet.
Crucially, the interaction with the $Z'$ gauge boson off-diagonally couples as
\begin{equation}
\mathcal{L}_\mathrm{int} =  g_{B-L} Z'{}^{\mu}\left( (\partial_{\mu}S) P- S \partial_{\mu}P \right) ,
\label{eq:Lag:gauge-DM-DM}
\end{equation}
 and similarly all generation quarks and leptons also interact to $Z'$ with corresponding charges.
The absence of $Z'$-DM-DM coupling means that the $Z'$-mediated DM scattering off
 with a nucleon is inelastic and ineffective for the mass splitting larger than
 the energy transfer in the scatterings~\cite{Hall:1997ah,TuckerSmith:2001hy}.
While we have investigated the regime where the mass splitting is too large
 for $Z'$-mediated scattering to be effective in Ref.~\cite{Okada:2019sbb},
 a moderate mass splitting allows $Z'$-mediated inelastic scattering off nucleons
 to take place, which plays a central role in our scenario.

\subsection{Thermal relic abundance}

We estimate the thermal relic abundance of the real scalar DM, $S$, by solving the Boltzmann equation,
\begin{equation}
 \frac{d n }{dt}+3H n =-\langle\sigma_\mathrm{eff} v\rangle ( n^2 - n_{\rm EQ}^2),
\label{eq:boltzman}
\end{equation}
 where $H$ and $n_{\rm EQ}$ are the Hubble parameter and the DM number density at thermal equilibrium,
 respectively~\cite{KolbTurner}.  
In our model, the main annihilation mode is coannihilation $S P \rightarrow f\bar{f}$ through $s$-channel $Z'$ exchange for $m_{Z'} > m_S$ and the annihilation mode $S S \rightarrow Z^\prime Z'$ by $u(t)$-channel $P$ exchange for $m_{Z'} < m_S$.
We use the effective thermal averaged annihilation cross section
\begin{equation}
 \langle\sigma_\mathrm{eff}v\rangle  = \sum_{i,j= S,P} \langle\sigma_{ij} v_{ij}\rangle\frac{n_i}{n_{\rm EQ}}\frac{n_j}{n_{\rm EQ}},
\end{equation}
 to include the coannihilation effects properly and $n$ in Eq.~(\ref{eq:boltzman}) should be understood as $n = \sum_i n_i$ for $i = S, P$~\cite{Griest:1990kh,Edsjo:1997bg}.

As is shown in Ref.~\cite{Okada:2019sbb},
 the right panel of Fig.~\ref{Fig:bldetection}  displays
 the relation between $m_S$ and $m_{Z^\prime}$ drawn with blue lines
 in order to reproduce the observed DM relic abundance $\Omega h^2\simeq 0.1$~\cite{Planck:2018vyg}.
The experimental bound on the mass of $Z^\prime$ in the $U(1)_{B-L}$ model has been derived based on
 the LEP and Tevarton data~\cite{Carena:2004xs} as well as
 the latest LHC Run 2 results~\cite{Amrith:2018yfb,Das:2019fee} that are expressed as a brown shaded region.
There are two cases reproducing the observed DM abundance:
 one is $m_S \sim$ several TeV by the resonant annihilation through $s$-channel $Z^\prime$ exchange and 
 the other is due to the $S S \rightarrow Z^\prime Z^\prime$ annihilation for $m_S > 10$ TeV.
For a smaller $g_{B-L}$, DM is overabundant and no solution to $\Omega h^2\simeq 0.1$ is found.
Since the former is only available for TeV scale WIMP interpretation of the LZ event,
 we find the mass spectum 
\begin{equation}
m_S \simeq \frac{1}{2}m_{Z^\prime}.
\label{eq:mass_specrtum}
\end{equation}

\subsection{Dark matter inelastic scattering with nuclei}

The spin-independent (SI) DM scattering with a nucleon $n$ takes place
 through $Z'$ exchange.
The effective Lagrangian
\begin{equation}
\mathcal{L}= \frac{g^2_{B-L}}{m^2_{Z'}} \left( (\partial_{\mu}S) P- S \partial_{\mu}P \right)\overline{n}\gamma^{\mu}n ,
\end{equation}
is obtained by integrating $Z'$.
The scattering cross section by the $Z^\prime$ mediation is given by~\cite{Jungman:1995df}
\begin{subequations}
\begin{align}
\sigma_{\rm SI} =& \frac{1}{64\pi}
 \left(\frac{m_n m_S}{m_n + m_S}\right)^2b_n^2 , \\
b_n^2 =&  \frac{4g_{B-L}^4}{m_{Z'}^4} = \frac{1}{4v_2^2}, 
\end{align}
\label{sigmaSI:scalar}%
\end{subequations}
 where the approximation $m_S \simeq m_P$ is used.
Notice that the magnitude of the cross section (\ref{sigmaSI:scalar})
 is a function of only the VEV of $B-L$ breaking $v_2$. 
On the other hand, the mass splitting $\sqrt{m_P^2-m_S^2} = \mathcal{O}(100)$ keV is important
 and sets a kinematic threshold that suppresses standard low-energy recoils, 
 shifting the peak of the differential event rate to higher recoil energies, as observed at LZ.

\begin{figure}[htbp] 
 \centering
    \begin{tabular}{c}
 \begin{minipage}{0.48\hsize}
\centering
\includegraphics[width=7.5cm]{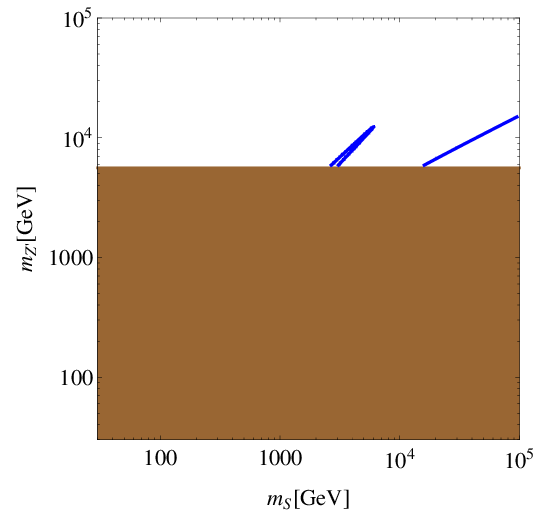}
 \end{minipage}
      \begin{minipage}{0.04\hsize}
        \hspace{2mm}
      \end{minipage}
 \begin{minipage}{0.46\hsize}
\centering
\includegraphics[width=7.5cm]{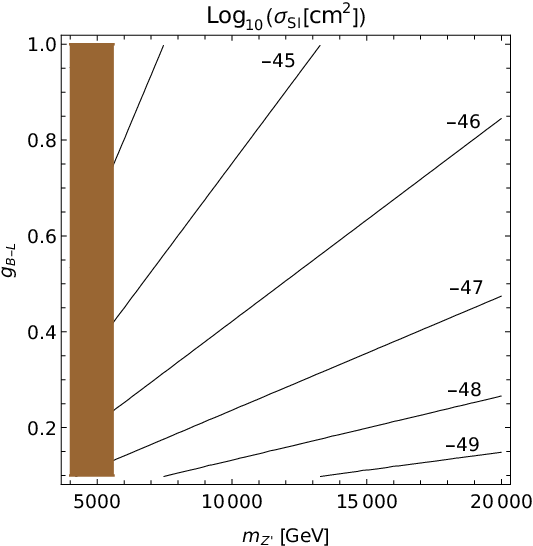}
 \end{minipage}   
 \end{tabular}
\caption{
The excluded region for the $Z'$ boson mass by the LHC is brown shaded.
Left panel: 
The contour (in blue) along which the observed DM relic abundance $\Omega h^2 \simeq 0.1$ is
 reproduced for $g_{B-L}=0.5$, taken from Ref.~\cite{Okada:2019sbb}.
Right panel: The contours of the WIMP-nucleon inelastic scattering cross section. 
}
\label{Fig:bldetection}
\end{figure}

We show in the right-pannel of Fig.~\ref{Fig:bldetection} the spin-independent cross section,
 $\sigma^{\mathrm{SI}}$.
From both panels, we find that for $m_S \simeq $ a few TeV and $g_{B-L}\sim 0.5$, 
the inelastic scattering cross section with a nucleon turns out to be
 of the order of $10^{-45}$ cm${}^2$,
 which perfectly matches the required rate for the inelastic DM interpretation
 of the LZ anomaly.

%

\section{Summary}
\label{Sec:Summary} 

We have demonstrated that the inelastic scalar DM model in
 a gauged $U(1)_{B-L}$ model can naturally account for the recent
 anomalous high-energy event reported by the LZ collaboration. 
The small mass splitting between DM states arises naturally
 through $B-L$ symmetry breaking and plays an important role
 to explain the kinematical property of the LZ event.
Our model is strongly restrictive.
To realize a thermal WIMP at the TeV scale,
 the mass relation (\ref{eq:mass_specrtum}) is required
 for the resonant annihilation to take place, and
 the $U(1)_{B-L}$ gauge coupling must be about $0.5$ to realize a sufficiently large annihilation cross section.
Thermal DM conditions indicate that the inelastic scattering cross-section with nucleons is of $10^{-45}$ cm${}^2$,
 and this magnitude is just right to explain the LZ event.
Because all key model parameters ($m_S$, $m_{Z'}$, $v_{B-L}$, and $g_{B-L}$) are tightly constrained
 by the combination of thermal relic and direct detection requirements,
 our scenario is highly predictive. 
It can be tested through $Z'$ boson searches at upcoming high-energy collider experiments.


\section*{Acknowledgments}
This work is supported in part by the U.S. DOE Grants No. DE-
SC0026347 (N.O.) and KAKENHI Grants No. JP23K03402 (O.S.).

%



\end{document}